\documentstyle[11pt]{article}

\begin{document} 

\makeatletter
\@addtoreset{equation}{section}
\makeatother
\renewcommand{\theequation}{\thesection.\arabic{equation}}
\baselineskip 15pt

\newtheorem{defofentang}{Definition}[section]
 \newtheorem{factorizability}{Theorem}[section]
 \newtheorem{rank}[factorizability]{Theorem}
 \newtheorem{onedimensional}[factorizability]{Theorem}
 \newtheorem{correlation}[factorizability]{Theorem}
 \newtheorem{defofentangidentical}{Definition}[section]
 \newtheorem{factorizabilityidentical}{Theorem}[section]
 \newtheorem{defofentangidentical2}[defofentangidentical]{Definition}
 \newtheorem{factorizabilityidentical2}[factorizabilityidentical]{Theorem}
 \newtheorem{defofentangidentical3}[defofentangidentical]{Theorem}
 \newtheorem{defofentangN}{Definition}[section]
 \newtheorem{onedimensionalN}{Theorem}[section]
 \newtheorem{factorizabilityN}[onedimensionalN]{Theorem}
 \newtheorem{rankN}[onedimensionalN]{Theorem}
 \newtheorem{corrbetween}[onedimensionalN]{Theorem}
 \newtheorem{mutualentang}[defofentangN]{Definition}
 \newtheorem{mutualfactor}[onedimensionalN]{Theorem}
 \newtheorem{sviluppodipi}{Theorem}[section]
 \newtheorem{giancarlo}[sviluppodipi]{Theorem}
 \newtheorem{definizionedient}{Definition}[section]
 \newtheorem{defentanggroupfermion}{Definition}[section]
 \newtheorem{defofentangNidentical}[defentanggroupfermion]{Definition}
 \newtheorem{relevant}[sviluppodipi]{Theorem}
 \newtheorem{teodianti}[sviluppodipi]{Theorem}

\title{\bf Entanglement and properties
\footnote{Work supported in part by Istituto Nazionale di Fisica Nucleare,
 Sezione di Trieste, Italy}}
\author{GianCarlo Ghirardi\footnote{e-mail: ghirardi@ts.infn.it}\\
{\small Department of Theoretical Physics of the University of Trieste, and}\\
{\small Istituto Nazionale di Fisica Nucleare, Sezione di Trieste, Italy and}\\
{\small International Centre for Theoretical Physics, Trieste, Italy}\\
and \\
\\ Luca Marinatto\footnote{e-mail: marinatto@ts.infn.it }\\
{\small Department of Theoretical Physics of the University of Trieste, and}\\
{\small Istituto Nazionale di Fisica Nucleare, Sezione di Trieste, Italy.}}

\date{}

\maketitle
\begin{abstract}
Various topics concerning the entanglement of composite quantum systems are
 considered with particular emphasis concerning the strict relations of such
 a problem with the one of attributing objective properties to the
 constituents. In particular we will focus our attention to composite
 quantum systems composed of identical constituents, with the purpose
 of dealing with subtle issues, which have never been adequately discussed 
 in the literature, originating from the true indistinguishability 
 of the subsystems involved.
\end{abstract}


\section{Introduction}

One of the crucial points of any theory aiming to account for natural
 phenomena concerns the possibility of identifying the properties objectively
 possessed by individual physical systems and/or by their constituents.
Such a problem acquires a completely different status within different
 theoretical schemes, typically in the classical and quantum cases.
First of all, quantum mechanics, if the completeness assumption is made,
 requires a radical change of attitude about the problem of attributing 
 objective properties to physical systems due to its fundamentally 
 probabilistic character.
Secondly, and even more important for our analysis, it gives rise to specific
 and puzzling situations concerning the properties of the constituents of a
 composite system due to its peculiar feature, Entanglement - the direct
 English translation of the original German form {\em Verschrankung} used by
 Schr\"odinger~\cite{ref1} - which Schr\"odinger himself considered
 {\em ``the characteristic trait of Quantum Mechanics, the one that enforces
 its entire departure from classical line of thoughts''}.
Quantum entanglement has played a central role in the historical development
 of quantum mechanics, since it has compelled the scientific
 community to face the essentially nonlocal features of natural processes.
Nowadays, entangled states have become the essential ingredients of all
 processes involving teleportation and quantum cryptography and constitute
 an important tool for implementing efficient quantum algorithms.
This explains why a great deal of efforts has been spent by theorists during
 the last years in trying to characterize the very nature and properties of
 entanglement, and this is also the reason which motivates our attempt to
 deepen some questions about these matters.


\section{Properties of individual physical systems}

Let us therefore start by discussing the notion of \textit{state of an
 individual physical system} within an hypothetical physical theory.
The crucial point, from a conceptual point of view, consists in identifying
 which is the most accurate characterization that the theory allows concerning
 the situation of such an individual physical system.
For our present purposes we assume that such a ``most exhaustive knowledge'' 
 is, in principle, possible and that it is expressed by mathematical entities 
 which we will denote as the {\bf States} of the theory (the so called
 {\em completeness assumption}).

One can immediately exhibit some elementary examples of what we have in mind.
For instance, the {\bf States} of a system of {\it N} point particles within
 Newtonian mechanics are the points $P$ of the $6N$-dimensional phase space
 of the system.
Within such a theory all conceivable observables, both
 referring to the whole system as well as to all its subsystems, are functions
 of the positions and momenta of the particles, so that, when one knows the
 {\bf State} of the system, i.e., the phase space point associated to it, one
 knows also the precise values of all physical observables.
Therefore we can claim that within Classical Mechanics all properties are 
 objectively possessed, in the precise sense that the measurement of any given 
 observable simply reveals the pre-existing value possessed by the
 observable.
It goes without saying that if we lack the complete information about the
 system, then we can make statements only concerning the (epistemic)
 probabilities that it possesses precise properties.
Nevertheless, it remains true that any individual system (and its subsystems)
 has all conceivable properties, in spite of the fact that we can be ignorant
 about them~\footnote{Stated differently, claims of the kind ``the energy of
 this particle has this specific value" have truth values, i.e. they are
 definitely either true or false.}.

As everybody knows, the situation is quite different in (non-relativistic)
 quantum mechanics where one asserts, with the completeness assumption, that 
 the {\bf States} of a system of $N$ particles are the
 state vectors $\vert \psi (1,\dots,N)\rangle$ of the associated Hilbert space.
Accordingly, the theory, in general, consents to make only nonepistemic
 probabilistic predictions about the outcomes of measurement processes even
 when the {\bf State} of the system is known.
However, in such a case, there are always complete sets of commuting
 observables such that the theory attaches probability one to a precise
 outcome in a measurement process of any one of them.
It is then natural to assume (as we will do) that when we can make certain
 (i.e. with probability one) predictions  about the outcomes, the system
 possesses objectively the property, or element of physical reality ``such
 an observable has such a value", independently of our decision to measure it.
Here we have used the expression {\em objective properties} and {\em elements
 of reality} with the same meaning that Einstein~\cite{ref2} gave them in
 the analysis of the EPR paradox$\,$:
\begin{quotation}
  {\sl If, without in any way disturbing a system, we can predict with
 certainty (i.e. with a probability equal to unity) the value of a physical
 quantity, then there exists an element of physical reality corresponding to
 this physical quantity.}
\end{quotation}
When one takes into account the just outlined situation, one can concisely
 express the lesson that quantum mechanics has taught us, by stating that
 within such a theory one cannot consider (even in principle) an individual
 physical system as possessing objectively too many properties.
Some of them can be legitimately considered as {\it actual}, all the other
 have the ontological status of \textit{potentialities}.
At any rate, according to the above remarks, a system in a
 pure state always has complete sets of definite and objective properties.

Up to this point we have confined our attention to quantum systems considered 
 as a whole.
However the phenomenon of quantum entanglement makes
 the situation much more puzzling when consideration is given to composite
 quantum systems and one raises the problem of the properties of their
 constituents.
As we will see, in such a case it is very common to meet situations (most
 of which arise as a consequence of the interactions between the constituents)
 in which the constituents themselves do not possess {\it any property
 whatsoever}.
This is a new feature which compels us to face a quite peculiar state of
 affairs: not only must one limit drastically the actual properties of
 physical systems (being in any case true that the system as a whole always
 has some properties), but one is forced also to accept that the parts of a
 composite system can have no property at all.
In this way  the quantum picture of the universe as an ``unbroken whole", 
 or as ``undivided", emerges.


\section{Entanglement of two distinguishable particles}

In this section we study the Entanglement between two distinguishable
 particles ${\cal S}_{1}$ and ${\cal S}_{2}$.
Let us suppose that the two particles are parts of a larger quantum system
 ${\cal S}={\cal S}_{1} + {\cal S}_{2}$, whose associated Hilbert space
 ${\cal H}$ is the direct product of the Hilbert spaces of the single
 subsystems, ${\cal H}={\cal H}_{1}\otimes {\cal H}_{2}$.
We will always assume that the
 composite quantum system ${\cal S}$ is  described by a state vector $\vert
 \psi(1,2) \rangle \in {\cal H}$ or, in a totally equivalent manner, by a
 pure density operator $\rho =\vert \psi(1,2) \rangle\langle \psi(1,2) \vert$.

Let us start by characterizing a non-entangled composite system
 by making explicit reference to the fact that one of its two constituent
 subsystems possesses complete sets of properties (as we will see this in
 turn implies that the same is true for the other constituent):
\begin{defofentang}
 \label{defofentang}
 The system ${\cal S}_{1}$, subsystem of a composite quantum system
 ${\cal S}={\cal S}_{1} + {\cal S}_{2}$ described by the pure density
 operator $\rho$, is {\bf non-entangled} with subsystem ${\cal S}_{2}$ if
 there exists a projection operator $P^{(1)}$ onto a  one-dimensional manifold
 of ${\cal H}_{1}$ such that:
\begin{equation}{\nonumber}
 \label{separabilitanonidentici}
  Tr^{(1+2)}[\,P^{(1)}\otimes I^{(2)}\rho\,]=1.
\end{equation}
\end{defofentang}
The fact that in the case of non-entangled states it is possible to consider
 each one of the constituents as possessing complete sets of well definite
 physical properties, independently of the existence of the other part,
 follows directly from the following theorem (for the proofs of the present
 and the following theorems see~\cite{gmw}):
\begin{factorizability}
  \label{factorizability}
  If consideration is given to a composite quantum system ${\cal S}=
 {\cal S}_{1} + {\cal S}_{2}$ described by the pure state vector $\vert
 \psi(1,2)\rangle $ (or, equivalently by the pure density operator
 $\rho =\vert \psi(1,2) \rangle\langle \psi(1,2) \vert$) of ${\cal H}=
 {\cal H}_{1}\otimes {\cal H}_{2}$, each of the following three conditions
 is necessary and sufficient in order that subsystem ${\cal S}_{1}$ is
 non-entangled with subsystem ${\cal S}_{2}$:
\begin{enumerate}
 \item  there exists a projection operator $P^{(1)}$ onto a one-dimensional
  manifold of ${\cal H}_{1}$ such that \\
  $Tr^{(1+2)}\left[ P^{(1)}\otimes I^{(2)}\rho \right] =1;$

 \item  the reduced statistical operator $\rho ^{(1)}=Tr^{(2)}\left[\,\rho\,
 \right] $ of subsystem $S_{1}$ is a  projection operator onto a
 one-dimensional manifold of ${\cal H}_{1}$;

 \item  the state vector $\left| \psi (1,2)\right\rangle $ is factorizable,
 i.e., there exist a state $\vert\phi(1) \rangle \in {\cal H}_{1}$ and a
 state $\vert \xi(2) \rangle \in {\cal H}_{2}$ such that $\vert \psi(1,2)
 \rangle= \vert \phi(1) \rangle\,\otimes \vert \xi(2) \rangle$.
\end{enumerate}
\end{factorizability}
Therefore if a quantum system composed of two subsystems is non-entangled, 
 the states of subsystems ${\cal S}_{1}$ and ${\cal S}_{2}$ are completely 
 specified, in the sense that it is possible to associate to each of them a 
 unique and well-defined state vector.
According to our previous discussion, the individual subsystems can therefore
 be thought of as having complete sets of definite and objective properties
 of their own.

We pass now to analyze the case of composite systems of two subsystems in
 entangled states. According to Theorem~\ref{factorizability}, the reduced
 density operator of each subsystem is not a projection operator onto a one
 dimensional manifold. It is then useful to analyze whether there exist
 projection operators on manifolds of dimension greater than or equal to 2 of
 ${\cal H}_{1}$, satisfying condition~(\ref{separabilitanonidentici}).
As shown by the following theorem, there
 is a strict relation between such projection operators and the range
 ${\cal R}[\, \rho^{(1)}\,]$ of the reduced statistical operator $\rho^{(1)}$:
\begin{rank}
 \label{rank}
 A necessary and sufficient condition for the  projection
 operator $P^{(1)}_{{{\cal{M}}_{1}}}$ onto the linear manifold
 ${\cal M}_{1}$ of ${\cal H}_{1}$ to satisfy the two following conditions:

\begin{enumerate}
 \item $Tr^{(1)}[\,P^{(1)}_{{\cal{M}}_{1}} \rho^{(1)}\,]=1$;
 \item there is no projection operator $\tilde{P}^{(1)}$ of ${\cal H}_{1}$
 satisfying the conditions $\tilde{P}^{(1)} < P^{(1)}_{{\cal{M}}_{1}}$
 ($\,$i.e. it projects onto a proper submanifold ${\cal N}_{1}$ of
 ${\cal M}_{1}$) and   $Tr^{(1)}[\, \tilde{P}^{(1)} \rho^{(1)}\,] =1$,
\end{enumerate}

\noindent is that the range ${\cal R}[\, \rho^{(1)}\,]$ of the reduced
 statistical operator $\rho^{(1)}$ coincides with ${\cal M}_{1}$.
\end{rank}
Let us analyze in detail the consequences of the above theorem by studying
 the following two cases concerning the range of the reduced statistical
 operator $\rho^{(1)}$:
\begin{enumerate}
 \item ${\cal R}[\,\rho^{(1)}\,]={\cal M}_{1} \subset {\cal H}_{1}$,
 \item ${\cal R}[\,\rho^{(1)}\,]={\cal H}_{1}$,
\end{enumerate}
In the first case, given any self-adjoint operator 
 $\Omega ^{(1)}$ of ${\cal H}_{1}$
 which commutes with $P^{(1)}_{{{\cal M}}_{1}},$ if consideration is given to
 the subset ${\cal B}$ (a Borel set) of its spectrum coinciding with the
 spectrum of its restriction $\Omega_{R}=P^{(1)}_{{{\cal M}}_{1}}\Omega^{(1)}
 P^{(1)}_{{{\cal M}}_{1}}$ to ${\cal M}_{1},$ we can state that subsystem
 ${\cal S}_{1}$ has the objective (in general unsharp) property that
 $\Omega^{(1)}$ has a value belonging to ${\cal B}$.
In particular, all operators which have ${\cal M}_{1}$ as an eigenmanifold,
 have a precise objective value.
Therefore, even though in the considered case we cannot say that subsystem
 ${\cal S}_{1}$ has a complete set of objective properties,
 it still has some sharp or unsharp properties associated to any observable
 which commutes with $P^{(1)}_{{{\cal M}}_{1}}$.

On the contrary, in the second case, we have to face the puzzling
 implications of entanglement in their full generality. 
In fact the only projection operator $P^{(1)}$ on ${\cal H}_{1}$ satisfying
 $Tr^{(1)}[P^{(1)}\rho ^{(1)}]=1$ is the identity operator $I^{(1)}$ on the
 Hilbert space ${\cal H}_{1}$.
The physical meaning of this fact amounts to state that subsystem 
 ${\cal S}_{1}$ does not possess objectively any
 sharp or unsharp property, i.e., that there is no self-adjoint operator for
 which one can claim with certainty that the outcome of its measurement will
 belong to any proper subset of its spectrum.

Another consequence of the entanglement of a composite quantum system is 
 the occurrence of strict correlations between observables of the
 component subsystems, even when they are far apart and non-interacting.
This is expressed by the following theorem:
\begin{correlation}
 \label{correlation}
 Subsystem ${\cal S}_{1}$ is non-entangled with subsystem ${\cal S}_{2}$
 iff, given the pure state $\vert \psi(1,2) \rangle$ of the composite system,
 the following equation holds for any pair of observables $A(1)$ of
 ${\cal H}_{1}$ and $B(2)$ of ${\cal H}_{2}$ such that $\vert\psi(1,2)\rangle$
 belongs to their domains:
\begin{equation}
 \label{correlazione}
   \langle \psi(1,2)\vert A(1)\otimes B(2) \vert\psi(1,2) \rangle =
   \langle \psi(1,2)\vert A(1)\otimes I^{(2)} \vert\psi(1,2) \rangle
   \langle \psi(1,2)\vert I^{(1)}\otimes B(2) \vert\psi(1,2) \rangle.
\end{equation}
\end{correlation}
The equation~(\ref{correlazione}) implies that no correlation exists
 between such pairs of observables.
 
Before passing to analyse the more interesting case of quantum systems composed
 of identical constituents, we outline that all the previous arguments, 
 definitions and theorems can be easily generalized to systems composed of 
 many distinguishable particles~\cite{gmw}.

 
\section{Entanglement of two identical particles}

The issue of attributing objective properties to the constituents of a 
 quantum system composed of identical particles, unfortunately does not turn 
 out to be a straightforward generalization of the just analysed case 
 involving distinguishable particles, and the problem of entanglement has to 
 be reconsidered. 
 
For example, the naive idea that the two systems being non-entangled requires
 and is guaranteed by the fact that their state vector is the direct product
 of vectors belonging to the corresponding Hilbert spaces, cannot be simply
 transposed to the case of interest.
One can easily realize that this must be the case by taking into account that
 the only allowed states for a system of two identical particles must exhibit
 precise symmetry properties under the exchange of the two particles.
If one would adopt the previous criterion one would be led to conclude
 (mistakenly) that non-entangled states of identical particles cannot
 exist. 
The inappropriateness of taking such a position derives from not taking into
 account various fundamental facts, in particular that identical particles are
 truly indistinguishable, so that one cannot pretend that a particular one
 of them has properties, and that the set of observables for such a system
 has to be restricted to the self-adjoint operators which are symmetric for
 the exchange of the variables referring to the two subsystems.
Accordingly, it goes without saying that, when dealing with the system, e.g.,
 of two electrons, we will never be interested in questions like ``is the
 electron which we have labeled $1$ at a certain position or is its spin
 aligned with a given axis?'' but our only concerns will be of the type: on
 the basis of the knowledge of the state vector describing the composite
 system, can one legitimately consider as objective a statement of the kind
 ``there is an electron in a certain region and it has its spin up along a
 considered direction''?

To prepare the ground for settling such issues, we begin by discussing the 
 case of two identical particles, linking the idea that they are non-entangled 
 to the request that both of them possess a complete set of properties.
Accordingly we give the following definition:
\begin{defofentangidentical}
 \label{defofentangidentical}
  The identical constituents ${\cal S}_{1}$ and ${\cal S}_{2}$ of a composite
  quantum system ${\cal S}={\cal S}_{1}+{\cal S}_{2}$ are {\bf non-entangled}
  when both constituents possess a complete set of properties.
\end{defofentangidentical}
Taking into account this fact we need to identify the conditions under which one

 can legitimately claim that one of the constituents possesses a complete 
 set of properties.
\begin{defofentangidentical2}
 \label{defofentangidentical2}
 Given a composite quantum system ${\cal S}={\cal S}_{1}+{\cal S}_{2}$ of
 two identical particles described by the pure density operator $\rho$, we
 will say that one of the constituents has a complete set of properties iff
 there exists a one dimensional projection operator $P$, defined on the Hilbert
 space ${\cal H}^{(1)}$ of each of the subsystems, such that:
\begin{equation}
\label{wittgenstein}
  Tr^{(1+2)}[\,E(1,2)\rho \,]=1
\end{equation}
\noindent where
\begin{equation}
  \label{operatorediproiezione}
   E(1,2)=P^{(1)}\otimes [\,I^{(2)}-P^{(2)}\,] + [\,I^{(1)}-P^{(1)}\,]\otimes
  P^{(2)} + P^{(1)}\otimes P^{(2)}.
\end{equation}
\end{defofentangidentical2}
The operator $E(1,2)$ is symmetric under the exchange of the labels of the two 
 particles and it is a projection operator.
Furthermore $Tr^{(1+2)}[\,E(1,2)\rho \,]$ gives the probability of finding
 {\em at least} one of the two identical particles in the state onto which the
 one-dimensional operator $P$ projects\footnote{We remark that one could drop
 the last  term in the expression~(\ref{operatorediproiezione}) getting an 
 operator whose
 expectation value would give the probability of precisely one particle having
 the properties associated to $P$. In the case of identical fermions this would 
 make no difference but for bosons it would not cover the case of both 
 particles  having precisely the same properties.}.

We are now able to link the fact that one constituent possesses a complete set 
 of properties to the explicit form of the state vector. 
This is specified by the following theorem:
\begin{factorizabilityidentical}
 \label{factorizabilityidentical}
 One of the identical constituents of a composite quantum system
 ${\cal S}={\cal S}_{1} + {\cal S}_{2}$, described by the pure normalized
 state $\vert \psi(1,2) \rangle$ has a complete set of properties
 iff $\vert \psi(1,2) \rangle$ is obtained by symmetrizing or antisymmetrizing
 a factorized state.
\end{factorizabilityidentical}
There follows that the process of symmetrization or antisymmetrization of a
 factorized quantum state describing a system composed of identical
 particles does not forbid to attribute a complete set of physical properties
 to the subsystems: the only claim that one cannot make is to attribute the
 possessed property to one rather than to the other constituent.
At this point it is straightforward to derive the following two theorems which
 give the desired sufficient and necessary conditions in order that a couple of 
 fermion and boson particles can be considered as non-entangled:
 \begin{factorizabilityidentical2}
 \label{factorizabilityidentical2}
The identical fermions ${\cal S}_{1}$ and ${\cal S}_{2}$ of a composite
 quantum system ${\cal S}={\cal S}_{1}+{\cal S}_{2}$ described by the pure
 normalized state $\vert \psi(1,2) \rangle$ are {\bf non-entangled} iff
 $\vert \psi(1,2) \rangle$ is obtained by antisymmetrizing a factorized state.
\end{factorizabilityidentical2}
\begin{defofentangidentical3}
 \label{defofentangidentical3}
  The identical bosons of a composite quantum system ${\cal S}={\cal S}_{1}
  + {\cal S}_{2}$ described by the pure normalized state $\vert \psi(1,2)
  \rangle$ are {\bf non-entangled} iff either the state is obtained by
  symmetrizing a factorized product of two orthogonal states or if it is the
  product of the same state for the two particles.
\end{defofentangidentical3}
%


\section{An illuminating example}

Let us consider a system of two identical spin 1/2 particles.
We notice that if one would confine his attention to the spin degrees of
 freedom alone, then, following our definitions and theorems, one would be led
 to conclude that the singlet state, which can be obtained by antisymmetrizing,
 e.g., the state $\vert z \uparrow \rangle_{1}\vert z\downarrow\rangle_{2},$
 would be a non-entangled state.
How does this fit with the general (and correct) position that such a state is, 
 in a sense, the paradigmatic case of an entangled two body system? 
The apparent contradiction can be easily solved by taking into account also 
 the position of the particles, in addition to the spin degrees of freedom. 
Let us for example consider the state which is obtained by antisymmetrizing 
 the factorized state $\vert z \uparrow \rangle_{1}\vert R \rangle_{1} 
 \vert z \downarrow \rangle_{2} \vert L \rangle_{2}$:
\begin{equation}
 \label{quisotto}
 \vert \psi(1,2)\rangle = \frac{1}{\sqrt{2}}\,[\,
 \vert z \uparrow \rangle_{1}\vert R \rangle_{1}
 \vert z \downarrow \rangle_{2} \vert L \rangle_{2} -
 \vert z \downarrow \rangle_{1}\vert L\rangle_{1}
 \vert z \uparrow \rangle_{2} \vert R \rangle_{2}\,]\:,
\end{equation}
where $\vert R \rangle$ and $\vert L \rangle$ are two orthogonal spatial 
 locations.
Even though for such a state it is meaningless to speak of particle $1$ as 
 distinguishable from 
 particle $2$, we can correctly state that {\em there is a particle with spin up
 along {\it z}-axis and located in region R} and that {\em there is a particle 
 with spin down and located in region L}.
Having attributed a complete set, i.e. position and spin, of properties to both 
 constituents, we can correctly state that the above state in non-entangled, 
 notwithstanding its non-factorized mathematical form. 

On the contrary, for a state like
\begin{equation}
\label{veroepr}
 \vert \psi(1,2)\rangle = \frac{1}{\sqrt{2}}\, [\, \vert z \uparrow \rangle_{1}
 \vert z \downarrow \rangle_{2} - \vert z \downarrow \rangle_{1}
 \vert z \uparrow \rangle_{2}\,] \otimes [\, \vert R \rangle_{1}
 \vert L \rangle_{2} + \vert L \rangle_{1}\vert R \rangle_{2}\,],
\end{equation}
which cannot be obtained by antisymmetrizing a factorized state, it is not
 possible, for example, to attribute any definite spin property to the particle
 located in $R$ and equivalently no definite spatial property can be attributed 
 to the particle with spin up.
In the case where $\vert R\rangle $ and $\vert L\rangle $ correspond
 to two distant spatial locations, the state vector of Eq.~(\ref{veroepr}) 
 represents the
 paradigmatic state considered in the usual EPR argument and in the
 experiments devised to reveal the non-local features of quantum mechanics.

The picture should now be clear: no state of two fermions in the
 singlet spin state can be obtained by antisymmetrizing a factorized wave
 function, when also the remaining degrees of freedom are taken into account.
In this sense one can understand how there is no contradiction between the usual
 statement that the singlet state is entangled and the fact that, if one
 disregards the spatial degrees of freedom, it can be obtained by
 antisymmetrizing a factorized spin state.


\section{Entanglement of $N$ indistinguishable particles}

Let us pass now to the case of $N$ indistinguishable particles and let us try 
 to define, in a conceptually correct way, the idea that the set of $N$ 
 identical particles we are dealing with, can be partitioned into 
 two ``subsets" of 
 cardinality $M$ and $K$, which are non-entangled with each other.
By following strictly the procedure we have introduced for the case of two
 particles, we will do this by first considering the possibility of 
 attributing a complete set of properties to each subset and we will give the 
 following definition:
\begin{definizionedient}
 \label{definizionedient}
 Given a quantum system of $N$ identical particles described by a pure
 state $\vert\psi^{(N)}\rangle$ we will say that it contains two 
 {\bf non-entangled}
 ``subgroups" of particles of cardinality $M$ and $K$, $(M+K=N)$, when both
 subgroups possess a complete set of properties.
\end{definizionedient}
In order to make statements of the sort we are interested in, i.e.,
 that objective properties can be attached to the subsets associated to a
 partition of the particles or, equivalently, that such subsets are
 non-entangled among themselves, we have to impose quite strict 
 constraints on the state vectors associated so such subsets (confining 
 for brevity our attention to the case of fermion 
 systems only~\footnote{For a more general and detailed analysis of all the 
 mathematical intricacies, which also includes the boson case, 
 we address the interested reader to the original paper~\cite{gmw}.}).
To this purpose we will begin by saying that the states 
 $\vert \Phi^{(K)} \rangle$ and 
 $\vert \Sigma^{(M)} \rangle$ belonging to ${\cal{H}}^{(K)}_{A}$ and 
 ${\cal{H}}^{(M)}_{A}$ (the antisymmetric manifolds of the Hilbert 
 spaces of $K$ and $M$ particles respectively) are 
 {\bf one-particle orthogonal} iff the following condition  holds
\begin{equation}
\label{int}
\int d\,X\,\Sigma^{(M)}(1,\dots,X,\dots,M)\,\Phi^{(K)\star}(1,\dots,X,\dots,K)
  = 0\:\: \:\:\:\:\:\forall \:X\:,
\end{equation} 
for every possible choice of the unsaturated variables. 
 
It is possible to prove that the above condition is equivalent to assuming
 the existence of a single particle basis 
 $\left\{ \vert \phi_{i} \rangle \right\}$ such that the Fourier expansion
 of $\vert \Phi^{(K)} \rangle$ involves only single particle states
 whose indices belong to $\Delta$, and $\vert \Sigma^{(M)}\rangle$
 involves only states whose indices belong to $\Delta^{\star}$, where $\Delta$ 
 and $\Delta^{\star}$ are two disjoint partition of the set of indices $i$.
For the sake of brevity we omit to describe the rather cumbersome procedure
 for identifying the generalization of the projection 
 operator of Eq.~(\ref{operatorediproiezione}), which allows us to speak 
 of properties
 objectively possessed by the considered subgroups of particles, and we pass 
 directly to stating the final interesting result:    
\begin{teodianti}
 \label{teodianti}
 Given a system $S^{(N)}$ of $N$ identical fermions in a pure state $\vert
 \psi^{(N)}\rangle$ of ${\cal{H}}_{A}^{(N)}$ it contains two non-entangled
 subsets of cardinality $M$ and $K$ iff  $\vert\psi^{(N)}\rangle$ can be
 written as:
\begin{equation}
 \vert \psi^{(N)} \rangle = \sqrt{ { N \choose K}} P_{A}\:[\: \vert
 \Pi^{(M)} (1,...,M) \rangle \otimes \vert \Phi^{(K)}(M+1,...,N) \rangle\: ]\:,
\end{equation}
where the states $\vert\Pi^{(M)}(1,...,M) \rangle$ and $\vert \Phi^{(K)}
 (M+1,...,N) \rangle$ are ``one-particle orthogonal" among themselves and
 $P_{A}$ is the projection operator onto the totally antisymmetric
 manifold ${\cal{H}}_{A}^{(N)}$.
\end{teodianti}
This theorem represents the correct generalization of the one already proved
 for the simpler case of two identical fermions, and once again it makes
 precise the mathematical conditions under which an arbitrary $N$-particle 
 state can (or cannot) describe entangled subgroups of particles.
It is worth noticing that, once more, the non-factorized form of the state
 required by the antisymmetrization postulate, does not imply by itself 
 the entanglement of the subsystems.


\section{One-particle orthogonality}
\label{thephysical}

Let us now discuss the physical motivations for the request of the 
 {\em one-particle orthogonality} in dealing with non-entangled subgroups of
 identical particles.
To this purpose, let us consider two closed submanifolds $V^{(M)}_{\Delta}$ and
 $V^{(K)}_{\Delta^{*}}$ of ${\cal H}^{(M)}_{A}$ and ${\cal H}^{(K)}_{A}$
 respectively, spanned by {\em one-particle orthogonal states}, as defined
 previously.
If $\{\vert\Upsilon^{(M)}_{j}\rangle\}$ and $\{\vert\Xi^{(K)}_{l}\rangle\}$
 are two orthonormal bases spanning such manifolds, given arbitrary 
 states $\vert\chi^{(M)}\rangle$ and $\vert\tau^{(M)}\rangle$ of
 $V^{(M)}_{\Delta}$, $\vert\mu^{(K)}\rangle$ and $\vert\nu^{(K)}\rangle$ of 
 $V^{(K)}_{\Delta^{*}}$, the following relations hold:
\begin{equation}
 \label{mio}
 \sum_{l}\left|\left[\sqrt{ { N \choose K}}\langle \chi^{(M)}\vert\langle
 \Xi^{(K)}_{l}
 \vert P_{A}\right]\cdotp \left[\sqrt{ { N \choose K}} P_{A} \vert
 \tau^{(M)}\rangle
 \vert \nu^{(K)}\rangle \right] \right|^{2}=\vert \langle \chi^{(M)} \vert
 \tau^{(M)}
 \rangle \vert^{2}\:,
\end{equation}
\noindent and
\begin{equation}
 \label{tuo}
 \sum_{j}\left \vert \left[ \sqrt{ {N \choose K}} \langle
 \Upsilon^{(M)}_{j}\vert\langle
 \mu^{(K)}\vert P_{A}\right]\cdotp \left[\sqrt{{ N \choose K}} P_{A} \vert
 \tau^{(M)}\rangle\vert\nu^{(K)}\rangle\right]\right|^{2}=\vert \langle
 \mu^{(K)} \vert
 \nu^{(K)} \rangle \vert^{2}\:.
\end{equation}
\noindent These equations show that, provided two one-particle orthogonal
 manifolds $V^{(M)}_{\Delta}$ and $V^{(K)}_{\Delta^{*}}$ can be identified,
 and provided the interactions between the particles determining the
 subsequent evolution do not alter the specific features of the state vector,
 then one can {\it do the physics within each manifold by disregarding the
 other one}, even though the appropriate antisymmetrization requests for the
  whole set of fermions are respected~\footnote{With reference to
 eq.(\ref{mio}), we stress that the one-particle orthogonality of the states
 $\vert\tau^{(M)}\rangle$ and $\vert\nu^{(K)}\rangle$, and
 $\vert\chi^{(M)}\rangle$ and $\vert\Xi^{(K)}\rangle$ as well as the
 corresponding ones for the states appearing in eq.~(\ref{tuo}), is absolutely
 fundamental - as the reader can check - in order that the (physically
 important) equality sign between the expressions at the left and right hand
 sides of the equations holds.}.
In fact scalar products between antisymmetrized products of one-particle 
 orthogonal states turns out to be equivalent to the evaluation of 
 {\em reduced} scalar products involving 
 only states of a certain subgroup of all the particles, neglecting the 
 presence of the others.

These considerations should have made clear the appropriateness of adopting
 our criteria for the attribution of complete sets
 of properties associated to $\vert\Pi ^{(M)}\rangle$ and
 $\vert\Phi ^{(K)}\rangle$ and for the identification of non-entangled subsets
 of a system of $N$ identical fermions.


\section{Some remarks concerning almost perfect non-entanglement}

Having made precise the idea of a ``group of particles'' of a system of
 identical particles being non-entangled with the remaining ones, we can
 analyse the following physical situation: there is a Helium atom 
 at the origin $O$ of our reference frame ({\em here}) and a Lithium atom 
 at a distance $d$ from $O$ ({\em there}).
Our worries concern the legitimacy of claiming ``there is a Helium atom at
 the origin'' or ``there is a Lithium atom at a distance $d$ from the origin"
 when one takes into account the identity of the electrons of the two
 systems which requires the state vector to be totally antisymmetric under
 their exchange~\footnote{To discuss the conceptually relevant aspects of this
 problem we will confine,
 for simplicity, our considerations only to the electrons which are involved,
 disregarding the nuclei of the atoms - and the necessary antisymmetrization
 concerning the protons and the neutrons.}.
In order to settle the issue let us pay attention to the total state
 vector of the complete system ``Helium+Lithium'':
  \begin{equation}
   \label{eliolitio}
   \vert \psi^{(5)} \rangle \propto {\cal{G}} \left[ \vert
  Helium^{(2)}here\rangle
   \otimes \vert Lithium^{(3)}there \rangle \right]\:,
  \end{equation}
where ${\cal G}$ is the permutation operator which exchange at least one 
 of the two particle indices of the Helium with those of the Lithium. 
  
Since the factors $\vert Helium^{(2)}here \rangle$
 and $\vert Lithium^{(3)}there\rangle $ do not exactly satisfy our fundamental 
 request of being one-particle orthogonal, due to the non-compactness of their
 spatial supports, the above state cannot be strictly considered as
 non-entangled.
However we can explicitly evaluate integrals like the one of
 Eq.~(\ref{int}), which, when they vanish, make legitimate precise claims
 concerning the objective properties possessed by the two subgroups of two and
 three electrons.
We notice that the modulus of such integrals is of the order of the
 overlap integrals of the electronic wave functions.
Since they decrease exponentially outside a region of $10^{-8}cm$ from the
 corresponding nuclei, we immediately see that, for a distance between the
 two nuclei of the order of $1cm,$ the relevant integral
 turns out to have a value of the order of $10^{-10^{16}}$.
So, strictly speaking, our claim ``there is a Helium atom around the origin'' 
 is not perfectly correct.
However, in the considered case, the formal conditions which are
 necessary for attributing consistently objective properties to ``a group of
 particles'' are satisfied to such an high degree of accuracy that 
 we can still {\em legitimately} say that ``there is a Helium atom here {\em
 and} 
 a Lithium atom there'' and that the whole state is (almost) non-entangled.


\section{Conclusions}

We have discussed entanglement from the physical point of view of its
 implications concerning the properties objectively possessed by subgroups 
 of constituents of a composed physical systems.
In the delicate case of physical systems involving identical particles 
 we have given consistent definitions and proved general theorems.
They should have clarified that the apparent and formal ``entanglement" arising
 from the (anti)symmetrization of the corresponding wave function does not 
 imply by itself, both from a conceptual and a practical point of view, 
 an actual form of entanglement.


 

\begin{thebibliography}{99}


 \bibitem{ref1} E.Schr\"odinger, {\it Naturwissenschaften}, {\bf 23}, 807
 (1935); English translation in: {\it Proc. Am. Philos. Soc.}, {\bf 124}, 323
 (1980).

 \bibitem{ref2} A.Einstein, B.Podolsky and N.Rosen,  {\it Phys. Rev.},
 {\bf 47}, 777 (1935).

 \bibitem{gmw} G.C.Ghirardi, L.Marinatto and T.Weber, {\it Journal of 
 Statistical Physics}, {\bf 108}, 49 (2002).


\end{thebibliography}
\end{document}